\documentstyle[prl,aps,epsf,rotate]{revtex}

\newcommand{\be}{\begin{eqnarray}\label}
\newcommand{\ee}{\end{eqnarray}}

\begin{document}
\twocolumn[\hsize\textwidth\columnwidth\hsize
     \csname @twocolumnfalse\endcsname

\title{Evidence for a dynamic origin of charge}
\author{W. A. Hofer}
\address{Dept. of Physics and Astronomy, University College London\\
Gower Street, London WC1E 6BT, E-mail: w.hofer@ucl.ac.uk}

\maketitle


\begin{abstract}

The fundamental equations of particle motion lead to a modified
Poisson equation including dynamic charge. This charge derives
from density oscillations of a particle; it is not discrete, but
continuous. Within the dynamic model of hydrogen it accounts for
all features of electron proton interactions, its origin are
density oscillations of the proton. We propose a new system of
electromagnetic units, based on meter, kilogram, and second,
bearing on these findings. The system has none of the
disadvantages of traditional three-unit systems. On the basis of
our theoretical model we can genuinely derive the scaling factor
between electromagnetic and mechanic variables, which is equal,
within a few percent, to Planck's constant h. The implications of
the results in view of unifying gravity and quantum theory are
discussed. It seems that the hypothetical solar gravity waves, in
the low frequency range of the electromagnetic spectrum, are open
to experimental detection.

\end{abstract}

\vspace{0.5cm}
{\it PACS:} 03.65.Bz, 03.70.+k, 04.60.-m, 11.10.-z

\vskip2pc]


\section{The nature of charge}
Since the discovery of the electron by J. J. Thomson
\cite{thomson97} the concept of electric charge has remained
nearly unchanged. Apart from Lorentz' extended electron
\cite{lorentz04}, or Abraham's electromagnetic electron
\cite{abraham03}, the charge of an electron remained a point like
entity, in one way or another related to electron mass
\cite{bender84,keller97}. In atomic nuclei we think of charge as a
smeared out region of space, which is structured by the elementary
constituents of nuclear particles, the quarks \cite{montan94}.

The first major modification in this picture occurred only in the
last decades, when experiments on the quantum hall effect
\cite{stormer82,hall90} suggested the existence of "fractional
charge" of electrons. Although this effect has later been
explained on the basis of standard theory \cite{laughlin83}, its
implications are worth a more thorough analysis. Because it cannot
be excluded that the same feature, fractional or even continuous
charge, will show up in other experiments, especially since
experimental practice more and more focuses on the properties of
single particles. And in this case the conventional picture, which
is based on discrete and unchangeable charge of particles, may
soon prove too narrow a frame of reference. It seems therefore
justified, at this point, to analyze the very nature of charge
itself. A nature, which would reveal itself as an answer to the
question: What is charge?

It must be noted, in this respect, that the picture of continuous
charge, in classical theories, is due to the omission of the
atomic structure of matter. In any modern sense, continuous charge
can only be recovered by considering dynamic processes within the
very particles themselves.

With this problem in mind, we reanalyze the fundamental equations
of intrinsic particle properties \cite{hofer98a}. The consequences
of this analysis are developed in two directions. First, we
determine the interface between mechanic and electromagnetic
properties of matter, where we find that only one fundamental
constant describes it: Planck's constant $\hbar$. And second, we
compute the fields of interaction within a hydrogen atom, where we
detect oscillations of the proton density of mass as their source.
Finally, the implications of our results in view of unifying
gravity and quantum theory are discussed and a new model of
gravity waves derived, which is open to experimental tests.

\section{The origin of dynamic charge}

The intrinsic vector field ${\bf E}({\bf r},t)$, the momentum
density ${\bf p}({\bf r},t)$, and the scalar field $\phi ({\bf
r},t)$ of a particle are described by (see \cite{hofer98a}, Eq.
(18)):

\be{1}
 {\bf E}({\bf r},t) = - \nabla \frac{1}{\bar{\sigma}} \,
\phi ({\bf r},t) +  \frac{1}{\bar{\sigma}} \,
\frac{\partial}{\partial t} \, {\bf p}({\bf r},t) \ee

Here $\bar{\sigma}$ is a dimensional constant introduced for
reasons of consistency. Rewriting the equation with the help of
the definitions:

\be{2}
\beta := \frac{1}{\bar{\sigma}} \qquad \beta \phi ({\bf
r},t) := \phi ({\bf r},t) \ee

we obtain the classical equation for the electric field, where in
place of a vector potential ${\bf A}({\bf r},t)$ we have the
momentum density ${\bf p}({\bf r},t)$. This similarity, as already
noticed, bears on the Lorentz gauge as an expression of the energy
principle (\cite{hofer98a} Eqs. (26) - (28)).

\be{3}
{\bf E}({\bf r},t) = - \nabla \, \phi ({\bf r},t) + \beta
\, \frac{\partial}{\partial t} \, {\bf p}({\bf r},t) \ee

Note that $\beta$ describes the interface between dynamic and
electromagnetic properties of the particle. Taking the gradient of
(\ref{3}) and using the continuity equation for ${\bf p}({\bf
r},t)$:

\be{4} \nabla \, {\bf p}({\bf r},t) +  \frac{\partial}{\partial t}
\rho ({\bf r},t) = 0 \ee

where $ \rho ({\bf r},t)$  is the density of mass, we get the
Poisson equation with an additional term. And if we include the
source equation for the electric field ${\bf E}({\bf r},t)$:

\be{5} \nabla \, {\bf E}({\bf r},t) = \sigma ({\bf r},t), \ee

$ \sigma ({\bf r},t) $ being the density of charge, $\epsilon$ set
to 1 for convenience, we end up with the modified Poisson
equation:

\be{6} \Delta \phi ({\bf r},t) = - \underbrace{\sigma ({\bf
r},t)}_{static \, charge} - \underbrace{\beta \,
\frac{\partial^2}{\partial t^2} \rho ({\bf r},t)}_{dynamic \,
charge} \ee

The first term in (\ref{6}) is the classical term in
electrostatics. The second term does not have a classical
analogue, it is an essentially novel source of the scalar field
$\phi$, its novelty is due to the fact, that no dynamic
interpretation of the vector potential $ {\bf A}({\bf r},t)$
exists, whereas, in the current framework, $ {\bf p}({\bf r},t)$
has a dynamic meaning: that of momentum density.

To appreciate the importance of the new term, think of an
aggregation of mass in a state of oscillation. In this case the
second derivative of $\rho$ is a periodic function, which is, by
virtue of Eq. (\ref{6}), equal to periodic charge. Then this
dynamic charge gives rise to a periodic scalar field $\phi$. This
field appears as a field of charge in periodic oscillations: hence
its name, dynamic charge. It should be noted that dynamic charge
is essentially different from a classical dipole: in that case the
field can {\it appear zero} (cancellation of opposing effects),
whereas in case of dynamic charge it {\it is zero}. Even, as shall
be seen presently, for monopole oscillations.

\section{Oscillations of a proton}

We demonstrate the implications of Eq. (\ref{6}) on an easy
example: the radial oscillations of a proton. The treatment is
confined to monopole oscillations, although the results can easily
be generalized to any multipole. Let a proton's radius be a
function of time, so that $r_{p} = r_{p}(t)$ will be:

\be{7} r_{p}(t) = R_{p} + d \cdot \sin \omega_{H} t \ee

Here $R_{p}$ is the original radius, $d$ the oscillation
amplitude, and $\omega_{H}$ its frequency. Then the volume of the
proton $V_{p}$ and, consequently, its density of mass $\rho_{p}$
depend on time. In first order approximation we get:

\be{8} \rho_{p}(t) &=& \frac{3 M_{p}}{4 \pi} \left(R_{p} + d \sin
\omega_{H} t \right)^{-3} \nonumber \\ \rho_{p}(t) &\approx&
\rho_{0} \left( 1 - x \sin \omega_{H} t \right) \qquad x :=
\frac{3 d}{R_{p}} \ee

The Poisson equation for the dynamic contribution to proton charge
then reads:

\be{9} \Delta \phi ({\bf r},t) = - \beta x \rho_{0} \omega_{H}^2
\sin \omega_{H} t \ee

Integrating over the volume of the proton we find for the dynamic
charge of the oscillating proton the expression:

\be{10} q_{D}(t) = \int_{V_{p}} d^3 r \beta x \rho_{0}
\omega_{H}^2 \sin \omega_{H} t = \beta x M_{p} \omega_{H}^2 \sin
\omega_{H} t \ee

This charge gives rise to a periodic field within the hydrogen
atom, as already analyzed in some detail  and in a slightly
different context \cite{hofer98b}. We shall turn to the
calculation of a hydrogen's fields of interaction in the following
sections. But in order to fully appreciate the meaning of the
dynamic aspect it is necessary to digress at this point and to
turn to the discussion of electromagnetic units.

\section{Natural electromagnetic units}

By virtue of the Poisson equation (\ref{6}) dynamic charge must be
dimensionally equal to static charge, which for a proton is + e.
But since it is, in the current framework, based on dynamic
variables, the choice of $\beta$ also defines the interface
between dynamic and electromagnetic units. From (\ref{10}) we get,
dimensionally:

\be{11} [e] = [\beta] [M_{p} \omega_{H}^2] \quad \Rightarrow \quad
[\beta] = \left[\frac{e}{M_{p} \omega_{H}^2}\right] \ee

The unit of $\beta$ is therefore, in SI units:

\be{12} [\beta] = C \cdot \frac{s^2}{kg} = C \cdot \frac{m^2}{J}
\qquad [SI] \ee

We define now the {\it natural system of electromagnetic units} by
setting $\beta$ equal to 1. Thus:

\be{13} [\beta] := 1 \qquad \Rightarrow \qquad [C] = \frac{J}{m^2}
\ee

The unit of charge C is then energy per unit area of a surface.
Why, it could be asked, should this definition make sense?
Because, would be the answer, it is the only suitable definition,
if electrostatic interactions are accomplished by photons.

Suppose a $\delta^3({\bf r} - {\bf r'})$ like region around ${\bf
r'}$ is the origin of photons interacting with another
$\delta^3({\bf r} - {\bf r''})$ like region around ${\bf r''}$.
Then ${\bf r'}$ is the location of charge. Due to the geometry of
the problem the interaction energy will decrease with the square
of $|{\bf r'} - {\bf r''}|$. What remains constant, and thus
characterizes the charge at ${\bf r'}$, is only the interaction
energy per surface unit. Thus the definition, which applies to all
$r^{-2}$ like interactions, also, in principle, to gravity.

Returning to the question of natural units, we find that all the
other electromagnetic units follow straightforward from the
fundamental equations \cite{hofer98a}. They are displayed in Table
\ref{table1}.

If we analyze the units in Lorentz' force equation, we observe, at
first glance, an inconsistency.

\be{22} {\bf F}_{L} = q \left({\bf E} + {\bf u} \times {\bf
B}\right) \ee

The unit on the left, Newton, is not equal to the unit on the
right. As a first step to solve the problem we include the
dielectric constant $\epsilon^{-1}$ in the equation, since this is
the conventional definition of the electric field ${\bf E}$. Then
we have:

\be{23} [{\bf F}_{L}] = \frac{N m}{m^2} \left(\frac{m^4}{N}
\frac{N}{m^3} + \frac{m}{s} \frac{N s}{m^4} \right) = N + N \cdot
\frac{N}{m^4} \ee

Interestingly, now the second term, which describes the magnetic
forces, is wrong in the same manner, the first term was before we
included the dielectric units. It seems thus, that the dimensional
problem can be solved by a constant $\eta$, which is dimensionally
equal to $\epsilon$, and by rewriting the force equation
(\ref{22}) in the following manner:

\be{24} {\bf F}_{L} = \frac{q}{\eta} \left({\bf E} + {\bf u}
\times {\bf B}\right) \ee \be{25} [\eta] = N m^{-4} = C m^{-3} =
[\sigma] \ee

The modification of (\ref{22}) has an implicit meaning, which is
worth being emphasized. It is common knowledge in special
relativity, that electric and magnetic fields are only different
aspects of a situation. They are part of a common field tensor
$F_{\mu \nu}$ and transform into each other by Lorentz
transformations. From this point of view the treatment of electric
and magnetic fields in the SI, where we end up with two different
constants ($\epsilon, \mu$), seems to go against the requirement
of simplicity. On the other hand, the approach in quantum field
theory, where one employs in general only a dimensionless constant
at the interface to electrodynamics, the finestructure constant
$\alpha$, is over the mark. Because the information, whether we
deal with the electromagnetic or the mechanic aspect of a
situation, is lost. The natural system, although not completely
free of difficulties, as seen further down, seems a suitable
compromise. Different aspects of the intrinsic properties, and
which are generally electromagnetic, are not distinguished, no
scaling is necessary between ${\bf p}, {\bf E}$ and ${\bf B}$. The
only constant necessary is at the interface to mechanic
properties, which is $\eta$. This also holds for the fields of
radiation, which we can describe by:

\be{26} \phi_{Rad}({\bf r},t) = \frac{1}{8 \pi \eta} \left({\bf
E}^2 + c^2 {\bf B}^2 \right) \ee

Note that in the natural system the usage, or the omission, of
$\eta$ ultimately determines, whether a variable is to be
interpreted as an electromagnetic or a mechanic property. Forces
and energies are mechanic, whereas momentum density is not. The
numerical value of $\eta$ has to be determined by explicit
calculations. This will be done in the next sections. We conclude
this section by comparing the natural system of electromagnetic
units to existing systems.

From an analysis of the Maxwell equations one finds three
dimensional constants $k_{1},k_{2},k_{3}$, and a dimensionless
one, $\alpha$, which acquire different values in different systems
(see Table \ref{table2}).

Judging by the number of dimensional constants it seems that the
natural system is most similar to the electrostatic system of
units. However, since we have defined a separate interface to
mechanic properties, it is free of the usual nuisance of
fractional exponents without a clear physical meaning. The other
difference is that c, in the esu, is a constant, whereas it only
signifies the velocity of a particle in the natural system. For
photons this velocity equals c, but for electrons it is generally
much smaller. We note in passing that all the fundamental
relations for the intrinsic fields remain valid. Also the
conventional relations for the forces of interaction and the
radiation energy remain functionally the same. Only the numerical
values will be different.

Comparing with existing systems we note three distinct advantages:
(i) The system reflects the dynamic origin of fields, and it is
based on only three fundamental units: m, kg, s. A separate
definition of the current is therefore obsolete. (ii) There is a
clear cut interface between mechanics (forces, energies), and
electrodynamics (fields of motion). (iii) The system provides a
common framework for macroscopic and microscopic processes.

\section{Interactions in hydrogen}

Returning to proton oscillations let us first restate the main
differences between a free electron and an electron in a hydrogen
atom \cite{hofer98b}: (i) The frequency of the hydrogen system is
constant $\omega_{H}$, as is the frequency of the electron wave.
It is thought to arise from the oscillation properties of a
proton. (ii) Due to this feature the wave equation of momentum
density ${\bf p}({\bf r},t)$ is not homogeneous, but
inhomogeneous:

\be{27} \Delta {\bf p}({\bf r},t) - \frac{1}{u^2}
\frac{\partial^2}{\partial t^2} {\bf p}({\bf r},t) = {\bf f}(t)
\delta^3 ({\bf r}) \ee

for a proton at ${\bf r} = 0$ of the coordinate system. The source
term is related to nuclear oscillations. We do not solve
(\ref{27}) directly, but use the energy principle to simplify the
problem. From a free electron it is known that the total intrinsic
energy density, the sum of a kinetic component $ \phi_{K}$ and a
field component $\phi_{EM}$ is a constant of motion
\cite{hofer98a}:

\be{28} \phi_{K}({\bf r}) + \phi_{EM}({\bf r}) = \rho_{0} u^2 \ee

where $u$ is the velocity of the electron and $\rho_{0}$ its
density amplitude. We adopt this notion of energy conservation
also for the hydrogen electron, we only modify it to account for
the spherical setup:

\be{29} \phi_{K}({\bf r}) + \phi_{EM}({\bf r}) =
\frac{\rho_{0}}{r^2} u^2 \ee

The radial velocity of the electron has discrete levels. Due to
the boundary values problem at the atomic radius, it depends on
the principal quantum number $n$. From the treatment of hydrogen
we recall for $u_{n}$ and $\rho_{0}$ the results \cite{hofer98b}:

\be{30} u_{n} = \frac{\omega_{H} R_{H}}{2 \pi n} \qquad \rho_{0} =
\frac{M_{e}}{2 \pi R_{H}} \ee

where $R_{H}$ is the radius of the hydrogen atom and $M_{e}$ the
mass of an electron. Since $\rho_{0}$ includes the kinetic as well
as the field components of electron ''mass'', e.g. in Eq.
(\ref{29}), we can define a momentum density ${\bf p}_{0}({\bf
r},t)$, which equally includes both components. As the velocity
$u_{n} = u_{n}(t)$ of the electron wave in hydrogen is periodic:

\be{31} {\bf u}_{n}(t) = u_{n} \cos \omega_{H} t \, {\bf e}^{r}
\ee

\noindent the momentum density ${\bf p}_{0}({\bf r},t)$ is given
by:

\be{32} {\bf p}_{0}({\bf r},t) = \frac{\rho_{0} u_{n}}{r^2} \cos
\omega_{H} t \, {\bf e}^{r} \ee

The combination of kinetic and field components in the variables
has a physical background: it bears on the result that photons
change both components of an electron wave \cite{hofer98a}. With
these definitions we can use the relation between the electric
field and the change of momentum, although now this equation
refers to both components:

\be{33} {\bf E}_{0}({\bf r},t) = \frac{\partial}{\partial t} {\bf
p}_{0}({\bf r},t) = - \frac{\rho_{0} u_{n}}{r^2} \omega_{H} \sin
\omega_{H} t \, {\bf e}^{r} \ee

Note that charge, by definition, is included in the electric field
itself. Integrating the dynamic charge of a proton from Eq.
(\ref{10}) and accounting for flow conservation in our spherical
setup, the field of a proton will be:

\be{34} {\bf E}_{0}({\bf r},t) = \frac{q_{D}}{r^2} = \frac{M_{p}
\omega_{H}^2}{r^2} x \, \sin \omega_{H} t \, {\bf e}^{r} \ee

Apart from a phase factor the two expressions must be equal.
Recalling the values of $u_{n}$ and $\rho_{0}$ from (\ref{30}),
the amplitude $x$ of proton oscillation can be computed. We
obtain:

\be{35} x = \frac{3 d}{R_{p}} = \frac{M_{e}}{(2 \pi)^2 M_{p}}
\cdot \frac{1}{n} \ee

In the highest state of excitation, which for the dynamic model is
$n = 1$, the amplitude is less than $10^{-5}$ times the proton
radius: Oscillations are therefore comparatively small. This
result indicates that the scale of energies within the proton is
much higher than within the electron, say. The result is therefore
well in keeping with existing nuclear models. For higher $n$, and
thus lower excitation energy, the amplitude becomes smaller and
vanishes for $n \rightarrow \, \infty$.

It is helpful to consider the different energy components within
the hydrogen atom at a single state, say $n = 1$, to understand,
how the electron is actually bound to the proton. The energy of
the electron consists of two components.

\be{36} \phi_{K}({\bf r},t) = \frac{\rho_{0} u_{1}^2}{r^2} \sin^2
k_{1} r \cos^2 \omega_{H} t \ee

is the kinetic component of electron energy ($k_{1}$ is now the
wavevector of the wave). As in the free case, the kinetic
component is accompanied by an intrinsic field, which accounts for
the energy principle (i.e. the requirement, that total energy
density at a given point is a constant of motion). Thus:

\be{37} \phi_{EM}({\bf r},t) = \frac{\rho_{0} u_{1}^2}{r^2} \cos^2
k_{1} r \cos^2 \omega_{H} t \ee

is the field component. The two components together make up for
the energy of the electron. Integrating over the volume of the
atom and a single period $\tau$ of the oscillation, we obtain:

\be{38} W_{el} &=& \frac{1}{\tau} \int_{0}^{\tau} dt \int_{V_{H}}
d^3 r \left(\phi_{K}({\bf r},t) + \phi_{EM}({\bf r},t) \right)
\nonumber \\ &=&
\frac{1}{2} M_{e} u_{1}^2 \ee

This is the energy of the electron in the hydrogen atom. $W_{el}$
is equal to 13.6 eV. The binding energy of the electron is the
{\it energy difference} between a free electron of velocity
$u_{1}$ and an electron in a hydrogen atom at the same velocity.
Since the energy of the free electron $W_{free}$ is:

\be{39} W_{free} = \hbar \omega_{H} = M_{e} u_{1}^2 \ee

\noindent the energy difference $\triangle W$ or the binding
energy comes to:

\be{40} \triangle W = W_{free} - W_{el} = \frac{1}{2} M_{e}
u_{1}^2 \ee

This value is also equal to 13.6 eV. It is, furthermore, the
energy contained in the photon field $\phi_{Rad}({\bf r},t) $ of
the proton's radiation

\be{41} W_{Rad} &=& \triangle W = \frac{1}{\tau} \int_{0}^{\tau} dt
\int_{V_{H}} d^3 r \, \phi_{Rad}({\bf r},t) \nonumber \\
&=& \frac{1}{2} M_{e}
u_{1}^2 \ee

This energy has to be gained by the electron in order to be freed
from its bond, it is the ionization energy of hydrogen. However,
in the dynamic picture  the electron is not thought to move as a
point particle in the static field of a central proton charge, the
electron is, in this model, a dynamic and oscillating structure,
which emits and absorbs energy constantly via the photon field of
the central proton. In a very limited sense, the picture is still
a statistical one, since the computation of energies involves the
average over  a full period.

\section{The meaning of $\eta$}

The last problem, we have to solve, is the determination of
$\eta$, the coupling constant  between electromagnetic and
mechanic variables. To this end we compute the energy of the
radiation field $W_{Rad}$, using Eqs. (\ref{26}), (\ref{34}), and
(\ref{35}). From (\ref{26}) and (\ref{34}) we obtain:

\be{42} \phi_{Rad}(r,t) = \frac{1}{8 \pi \eta} {\bf E}^2 =
\frac{1}{8 \pi \eta} \cdot \frac{M_{p}^2 \omega_{H}^4}{r^4} x^2
\sin^2 \omega_{H} t \ee

Integrating over one period and the volume of the atom this gives:

\be{43} W_{Rad} &=& \frac{1}{\tau} \int_{0}^{\tau} dt
\int_{R_{p}}^{R_{H}} 4 \pi r^2 dr \, \phi_{Rad}({\bf r},t)
\nonumber \\ &\approx&
- \frac{1}{4 \eta} \cdot \frac{M_{p}^2 \omega_{H}^4 x^2}{R_{p}}
\ee

provided $R_{p}$, the radius of the proton is much smaller than
the radius of the atom. With the help of (\ref{35}), and
remembering that $W_{Rad}$ for $n = 1$ equals half the electron's
free energy $\hbar \omega_{H}$, this finally leads to:

\be{44} W_{Rad} = \frac{1}{4 \eta} \cdot \frac{M_{p}^2
\omega_{H}^4 x^2}{R_{p}} = \frac{1}{2} \hbar \omega_{H} \ee

\be{45} \eta = \frac{M_{e}^2 \nu_{H}^3}{2 h R_{p}} = \frac{1.78
\times 10^{20}}{R_{p}} \ee

since the frequency $ \nu_{H}$ of the hydrogen atom equals $6.57
\times 10^{15}$ Hz. Then $\eta$ can be calculated in terms of the
proton radius $R_{p}$. This radius has to be inferred from
experimental data, the currently most likely parametrization being
\cite{eisberg85}:

\be{46} \frac{\rho_{p}(r)}{\rho_{p,0}} = \frac{1}{1 + e^{(r -
1.07)/0.55}} \ee

radii in fm. If the radius of a proton is defined as the radius,
where the density $\rho_{p,0}$ has decreased to $\rho_{p,0}/e$,
with e the Euler number, then the value is between 1.3 and 1.4 fm.
Computing $4 \pi$ the inverse of $\eta$, we get, numerically:

\begin{eqnarray} \frac{4 \pi}{\eta} &=& 0.92 \times 10^{-34} \qquad (R_{p} =
1.3 fm) \nonumber \\ &=& 0.99 \times 10^{-34} \qquad (R_{p} = 1.4
fm)
\\ & = & 1.06 \times 10^{- 34} \qquad (R_{p} = 1.5 fm) \nonumber \ee

Numerically, this value is equal to the numerical value of
Planck's constant $\hbar$ \cite{uip78}:

\be{47} \hbar_{UIP} = 1.0546 \times 10^{-34} \ee

Given the conceptual difference in computing the radius the
agreement seems remarkable. Note that this is a genuine derivation
of $\hbar$, because nuclear forces and radii fall completely
outside the scope of the theory in its present form. If
measurements of $R_{p}$ were any different, then we would be
faced, at this point, with a meaningless numerical value.
Reversing the argument it can be said, that the correct value - or
rather the meaningful value - is a strong argument for the
correctness of our theoretical assumptions. Since these
assumptions involve to a greater or lesser extent the whole theory
of matter waves developed so far, we devote the rest of this
section mainly to a critical analysis of this result and shall
show the most striking physical implications only at the end.

Starting with the approximations involved, we note (i) a first
order approximation in d, and (ii) an approximation in the
integration. Since $d \approx 10^{-5} R_{p}$, and $R_{p} \approx
10^{-5} R_{H}$, both errors are negligible. In view of the
standard experimental error margins, also the deviation of a few
percent, depending on how we define the proton radius, seems
acceptable. On the positive side, there are two plausibility
arguments, indicating that we deal not only with a numerical
coincidence: (i) The Planck constant describes the interface
between frequency and energy in all fundamental experiments. Since
we started with a frequency (= proton oscillations), and
calculated an energy, it must have, at some point, entered the
calculation. The only unknown quantity in the calculation was
$\eta$: therefore it should contain $\hbar$. (ii) What we have in
fact developed with this model of hydrogen, is in spirit very
close to the harmonic oscillator in quantum theory; the rest
energy term is related to the energy of our photon field. In order
to be compatible with quantum theory, the energy must contain
$\hbar$. Again, the only variable, which could contain it, is
$\eta$.

It can also be asked, why electromagnetic variables are multiplied
by $\hbar$ to give the energy of radiation. Especially, since the
finestructure constant contains a division by $\hbar$:

\be{48} \alpha = \frac{e^2}{\hbar c} \ee

To answer this question, consider a variable in electrodynamics
$A_{ED}$ and its correlating variable $A_{M}$ in mechanics. Then
the transition from $A_{ED}$ to $A_{M}$ is described by a
transformation $T$, so that:

\be{49} A_{M} = T_{M,ED} A_{ED} \ee

Since the inverse transformation must exist and the variables are
assumed to be unique, the transformation is unitary:

\be{50} T_{M,ED} T_{ED,M} = T_{M,ED} T_{M,ED}^{-1} = 1 \ee

In our case the primary variables are the electromagnetic ones:
${\bf p}, {\bf E}, {\bf B}$. And the transformation involves a
multiplication by $\hbar$.

\be{51} A_{M} = \hbar A_{ED} \ee

The fundamental units m, kg, s are, in this system, the natural
system, tied to the electromagnetic variables. In quantum theory,
on the other hand, the fundamental variables are Newtonian. Then
the transformation between electromagnetic and mechanic variables
involves the inverse transformation.

\be{52} A_{M}(QM) = \hbar^{-1} A_{ED}(QM) \ee

If we consider, in addition, that charge has been included in the
definition of ${\bf E}$, the transformation, in conventional units
and in quantum theory should read:

\be{53} A_{M}(QM) = \frac{e^2}{\hbar} A_{ED}(QM) \ee

which is the finestructure constant multiplied by $c$. And $c$ is,
generally, only a matter of convention. Therefore we think, the
conclusion, that $4 \pi / \eta$ really is $\hbar$, is a reasonable
and safe one. But in this case Planck's constant has not much
bearing on a different scale of measurement, as is often invoked,
when there is talk about $\hbar \rightarrow 0$ in the macroscopic
scale. The constant bears on two fundamental aspects of matter
itself. As we see it, $\hbar$ describes the interface between
electromagnetic and mechanic variables of matter. It is therefore
even more fundamental than currently assumed. For the following,
let us redefine the symbol $\hbar$  by:

\be{54} \hbar := 1.0546 \times 10^{-34} [N^{-1} m^4] \ee

Then we can rewrite the equations for ${\bf F}$, the Lorentz
force, for ${\bf L}$, angular momentum related to this force, and
$\phi_{Rad}$, the radiation energy density of a photon in a very
suggestive form:

\be{55} {\bf F} = \hbar q \left(\frac{\bf E}{4 \pi} + {\bf u}
\times \frac{\bf B}{4 \pi} \right) \ee

\be{56} {\bf L} = \hbar q \,{\bf r} \times \left(\frac{\bf E}{4
\pi} + {\bf u} \times \frac{\bf B}{4 \pi} \right) \ee

\be{57} \phi_{Rad} = \frac{\hbar}{2} \left[\left(\frac{\bf E}{4
\pi}\right)^2 + c^2 \left(\frac{\bf B}{4 \pi}\right)^2 \right] \ee

Every calculation of mechanic properties involves a multiplication
by $\hbar$. Since $\hbar$ is a scaling constant, the term
''quantization'', commonly used in this context, is misleading.
Furthermore, it is completely irrelevant, whether we compute an
integral property (the force in (\ref{55})), or a density
($\phi_{Rad}$ in (\ref{57}), a force density can also be obtained
by replacing charge q by a density value). From the interaction
fields within a hydrogen atom, e.g. Eq. (\ref{42}), it is clear
that the field varies locally and temporary and can reach any
value between zero and its maximum. Although it is described by:

\be{58} \phi_{Rad}(r,t) = \frac{\hbar}{2} \frac{M_{p}^2
\omega_{H}^4}{r^4} x^2 \sin^2 \omega_{H}t \ee

it is not ''quantized''. Neither would be the forces based on the
field ${\bf E}_{0}$, or the angular momenta. Although, in both
cases, they are proportional to $\hbar$. What is, in a sense,
discontinuous, is the mass contained in the shell of the atom. But
this mass depends, as does the amplitude of $\phi_{Rad}(r,t)$, on
the mass of the atomic nucleus. So the only discontinuity left on
the fundamental level, is the mass of atomic nuclei. That the
energy spectrum of atoms is discrete, is a trivial observation in
view of boundary conditions and finite radii. To sum up the
argument: {\em There are no quantum jumps.}

All our calculations so far focus on single atoms. To get the
values of mechanic variables in SI units used in macrophysics, we
have to include the scaling between the atomic domain and the
domain of everyday measurements. Without proof, we assume this
value to be $N_{A}$, Avogadro's number. The scale can be made
plausible from solid state physics, where statistics on the
properties of single electrons generally involve a number of
$N_{A}$ particles in a volume of unit dimensions
\cite{ashcroft74}. And a dimensionless constant does not show up
in any dimensional analysis.

\section{Solar gravity fields}

We conclude this paper, which seems to open a new perspective on a
number of very fundamental problems, by a brief discussion. The
first issue concerns the nature of gravity. From the given
treatment it is possible to conclude, that there is maybe no
fundamental difference between electrostatic and gravitational
interaction. Both seem to be transmitted with the velocity of
light, both obey a $r^{-2}$ law, both are related to the existence
of mass, whether its static or its dynamic features. So the
conjecture, that also gravity is transmitted by a ''photon'', has
it least some basis. But here the similarities end. Because of the
vast differences in the coupling $\epsilon G \approx 10^{-22}$ one
must assume a very different frequency scale. From Eq. (\ref{34}):

\be{59} |{\bf E}| = \frac{q_{D}}{r^2} \approx \frac{M
\omega^2}{r^2} \ee

it can be inferred that the frequency scale for gravity and
electrostatics would differ by about $10^{-11}$.

\be{60} \omega_{G} \approx 10^{-11} \omega_{E} \ee

Here $\omega_{E}$ is the characteristic electromagnetic frequency,
$\omega_{G}$ its gravitational counterpart. The hypothesis can in
principle be tested. If we assume that $\nu_{G}$, the frequency of
gravity radiation, is about $10^{-11}$ times the frequency of
proton oscillation, we get:

\be{61} \nu_{G} \approx 1 - 100 \quad kHz \ee

If therefore electromagnetic fields of this frequency range exist
in space, we would attribute these fields to solar gravity. To
estimate the intensity of these, hypothetical, waves, we use Eq.
(\ref{33}):

\be{62} {\bf G}_{S}({\bf r},t) = \frac{\partial}{\partial t} {\bf
p}_{E}({\bf r},t) \ee

Here ${\bf G}_{S}$ is the solar gravity field. The momentum
density and its derivative can be inferred from centrifugal
acceleration.

\be{63} \frac{\partial}{\partial t} \, {\bf p}_{E}({\bf r},t) =
\rho_{E} \, a_{C} \, {\bf e}^{r} \nonumber \\ \rho_{E} = \frac{3
M_{E}}{4 \pi R_{E}^3} \qquad a_{C} = \omega_{E}^2 \, R_{O} \ee

where $R_{O}$ is the earth's orbital radius and where we have
assumed isotropic distribution of terrestrial mass. Then Eq.
(\ref{42}) leads to:

\be{64} \phi_{G}(r = R_{O}) = \frac{\hbar}{2} \left(
\frac{G_{S}}{4 \pi} \right)^2 = \frac{\hbar}{2} \left( \frac{3
M_{E} R_{O}}{4 R_{E}^3 \tau_{E}^2} \right)^2 \ee

Note the occurrence of Planck's constant also in this equation,
although all masses and distances are astronomical. The intensity
of the field, if calculated from (\ref{64}), is very small. To
give it in common measures, we compute the flow of gravitational
energy through a surface element at the earth's position. In SI
units we get:

\be{65} J_{G} (R_{O}) = \phi_{G}(R_{O}) \cdot N_{A} \cdot c
\approx 70 mW/m^2 \ee

Compared to radiation in the near visible range - the solar
radiation amounts to over 300 Watt/m$^2$ \cite{morrison96} - the
value seems rather small. But considering, that also radiation in
the visible range could have an impact on terrestrial motion, the
intensity of the gravity waves could be, in fact, much higher.

\section{Is there static charge?}

In the conventional models a particle's charge is not only
discrete, but has also a defined sign. Although anti-particles are
thought to exist, the charge of protons is positive, the charge of
electrons negative. Dynamic charge is neither discrete, nor does
it possess a defined sign. Depending on the exact moment, the
charge of a proton:

\be{66} q_{p} = M_{p} \omega_{H}^2 x \,\sin \omega_{H} t \ee

either has a positive or a negative value, which determines the
direction of the energy flow within the hydrogen atom. The
difference between electrons and protons in this model is mainly
due to their density of mass.

Related to this feature is a shift of focus within the dynamic
model of atoms. Although the states of the atom are described by
quantum numbers ($n$ for the principal state, $l$ and $m$ if
multipoles are included), these numbers refer primarily to nuclear
states of oscillation. States of the atom's electron are merely a
reaction to them. Therefore the properties of an atom, in the
dynamic model, refer to properties of the atomic nucleus. How this
model bears on chemical properties, remains to be seen.

The last issue is a consequence of our treatment of the hydrogen
atom. In this case the main features, the energy spectrum as well
as ionization energy and the energy of emitted photons can be
explained from dynamic charge alone. There is, in contrast to the
conventional treatment, no necessity to invoke static potentials.
It will also have been noted that in natural units and based on
dynamic processes interactions are generally free of any notion of
''charge'' in its proper sense. So does that mean, it could be
asked, that there is no charge? Based on the current evidence and
considering the situation in high energy physics, this is
definitely too bold a statement. Considering, though, that the
notion of a fixed ''elementary charge'' lies at the heart of all
current accounts of these experiments, the degree of theoretical
freedom in the dynamic picture is incomparably higher. So that
still, after a few years of development, we might end up with a
tentative answer: maybe not. And in this case, the question about
the true nature of charge will have been answered. It is dynamic
in nature, we would then say.

\section{Conclusions}

In this paper we presented evidence for the existence of a dynamic
component of charge. It derives, as shown, from the variation of a
particle's density of mass. A new system of electromagnetic units,
the natural system, has been developed, which bears on these
dynamic sources. We have given a fully deterministic treatment of
hydrogen, where we used our theoretical model to determine the
fundamental scaling constant between electromagnetic and mechanic
variables. The constant, we found, is $\hbar$, Planck's constant.
The constant thus has no bearing on any length scale, as
frequently thought. And finally we have discussed these results in
view of unifying gravity and quantum theory. The intensity of the
postulated solar gravity waves seems sufficiently high, so that
these waves, in the low frequency range of the electromagnetic
spectrum, can in principle be detected.

\section*{Acknowledgements}

I'd like to thank Jaime Keller. In our discussions I realized, for
the first time, that the most efficient theory of electrodynamics
might be one without electrons.



\begin{table}
\begin{tabular}{lll}
  Quantity & Symbol & Units \\
  & &  \\
  Charge   & $C_{SI}$ & $J m^{-2}$  \\
  Ampere   & $A_{SI}$ & $J m^{-2} s^{-1}$ \\
  & & \\
  Current density &  ${\bf J} $ & $J m^{-2} s^{-1}  m^{-2}$ \\
  Electric field &  ${\bf E}$  &   $N m^{-3}$ \\
  Magnetic field & ${\bf B}$  & $N s m^{-1} m^{-3}$ \\
  Scalar potential & $ \phi $ & $J  m^{-3}$ \\
  & & \\
  Dielectric constant & $\epsilon$ & $N m^{-4} = C m^{-3}$ \\
  Magnetic permeability & $\mu$ & $(C m^{-3})^{-1} m^{2} s^{-2}$
  \\
\end{tabular}
  \vspace{0.5 cm}
  \centering
  \caption{Electromagnetic quantities in natural units}\label{table1}
\end{table}

\begin{table}
\begin{tabular}{lcccc}
  System of units & $k_{1}$ & $k_{2}$ & $k_{3}$ & $\alpha$ \\
   & & &  & \\
  Electrostatic (esu) & 1   & $c^{-2}$ & 1 & 1     \\
  Electromagnetic     & $c^2$ & 1      & 1 & 1     \\
  Gaussian            & 1   & $c^{-2}$ & $c$ & $c^{-1}$ \\
  Heavyside-Lorentz   & $1/4 \pi$ & $1/4 \pi c^2$ & $c$ & $c^{-1}$ \\
  SI                  & $1/4 \pi \epsilon$ & $\mu/4 \pi$ & 1 & 1 \\
  Natural system      & $1/4 \pi$ & $c^{-2}$ & 1 & $1/4 \pi$
\end{tabular}
  \vspace{0.5 cm}
  \centering
  \caption{Magnitude and dimension of the electromagnetic constants.
           Note that we have taken the constants as they appear in
           the Maxwell equations (Eq. (A8) of [12]).}
           \label{table2}
\end{table}

\end{document}